\documentclass[twocolumn,aps,groupedaddress]{revtex4}

\usepackage{graphicx}
\usepackage{dcolumn}
\usepackage{bm}
\usepackage{amsmath,amssymb,amsfonts}
\usepackage{float}
\usepackage{color}
\usepackage{dcolumn}
\usepackage{wasysym}
\newcolumntype{N}{@{}m{0pt}@{}}

\begin{document}

\title{Density functional approach to correlated moir\'e states:
itinerant magnetism}

\author{Yang Zhang}
\author{Hiroki Isobe}
\author{Liang Fu}
\affiliation{Department of Physics, Massachusetts Institute of Technology, Cambridge, Massachusetts 02139, USA}
\begin{abstract}
Two-dimensional moir\'e superlattices have recently emerged as a fertile ground for creating novel electronic phases of matter with unprecedented control. Despite intensive efforts,  theoretical investigation of correlated moir\'e systems has been challenged by the large number of atoms in a superlattice unit cell  and the inherent difficulty of treating electron correlation. The physics of correlated moir\'e systems is governed by low-energy electrons in a coarse-grained  long-wavelength potential, unlike the singular Coulomb potential of atomically-spaced ions in natural solids.
Motivated by the separation between moir\'e and atomic length scales, in this work we apply density functional theory to study directly the continuum model of interacting electrons in the periodic moir\'e potential. Using this quantitatively accurate method, we predict itinerant spin-valley ferromagnetism in transition metal dichalchogenide heterobilayers, which originates from the constructive interplay between moir\'e potential and Coulomb interaction in a two-dimensional electron system.
\end{abstract}

\maketitle

Recent experiments on twisted bilayer graphene (TBG) \cite{cao2018correlated,cao2018unconventional,lu2019superconductors,kerelsky2019maximized,jiang2019charge,xie2019spectroscopic,choi2019electronic,yankowitz2019tuning,codecido2019correlated,sharpe2019emergent,tomarken2019electronic,zondiner2019cascade}, graphene-hBN heterostructures \cite{chen2019evidence,chen2019sig,serlin2019intrinsic,liu2019spin} and bilayer transition metal dichalcogenides (TMD) \cite{tang2020simulation,regan2020mott,wang2019magic} uncovered a whole new world of strongly interacting electrons in moir\'e superlattices, forming a variety of correlated states such as unconventional superconductors and Chern insulators.
From the theoretical point of view,
an important task is to develop a systematic and quantitatively accurate method for tackling correlated electron states in moir\'e materials.

 Despite the complexity associated with a large moir\'e unit cell containing  $\sim 10,000$ atoms, the physics of correlated states in semimetal- or semiconductor-based
 moir\'e materials is governed by a small fraction of electrons (holes) near the conduction (valence) band edge. Here the revelent length scale  far exceeds the interatmoic distance, which justifies a continuum description based on effective field theory. A celebrated example is the non-interacting continuum model for TBG, which leads to the prediction of flat bands at the magic twist angle \cite{bistritzer2011moire}.

To treat electron interaction within the continuum theory, the Hartree--Fock (HF) approximation has been widely employed \cite{guinea2018electrostatic, cea2019electronic, xie2020nature, bultinck2019ground,liu2019correlated,zhang2020correlated}. 
While it has yielded important insights on the insulating states in magic-angle graphene, the HF approximation often produces large errors in the ground-state energy and overestimate the tendency towards ferromagnetism. Moreover, the HF method is known to be ill-suited for treating metals, which are inherently unstable within this formalism. The lack of a reliable theoretical method hinders the study of  correlated metallic states in moir\'e materials at generic fillings.

In this work, we introduce a density-functional-theory (DFT) based method for tackling correlated moir\'e states. Rather than including all the electrons of $\sim 10,000$ atoms in the moir\'e unit cell, 
we directly target low-energy electrons/holes under long-wavelength moir\'e potential, which form a two-dimensional Coulomb system described by the interacting continuum model. Our method combines the conceptual simplicity of continuum theory and the quantitative accuracy of DFT. It is predictive, efficient and versatile.


Applying this ``moir\'e-DFT'' method to TMD heterobilayers such as WSe$_2$/WS$_2$, we find that at large moir\'e period,  the ground state in a range of fillings around $n=1.3$ holes per unit cell is an itinerant ferromagnet with full spin-valley polarization. By tracking the ground state energy as a function of the moir\'e potential strength and the filling, we show that the itinerant ferromagnetism originates from the constructive interplay between Coulomb interaction and the periodic moir\'e potential, which goes beyond the scope of single-band Hubbard model.

Our results shed light on a recent optical experiment on hole-doped WSe$_2$/WS$_2$ suggesting a possible transition between distinct ground-state magnetic orders around the filling $n=1.2$ \cite{tang2020simulation}.   
Based on moir\'e-DFT calculations, we find the critical twist angle for the onset of itinerant ferromagnetism, and propose transport and spectroscopic experiments for its detection.

Monolayer TMDs such as WSe$_2$, WS$_2$, MoS$_2$ and MoSe$_2$ are semiconductors with a direct band gap located at $K$ and $K'$ points of the Brillouin zone.
Due to spin-valley locking, holes at $K$ and $K'$ valleys carry opposite out-of-plane spins denoted by $s_z=\uparrow, \downarrow$. The energy dispersions of holes in the two valleys are identical within the effective mass approximation. In a TMD heterobilayer WSe$_2$/WS$_2$,
holes in the WSe$_2$ layer experience a long-wavelength periodic potential $V(\mathbf{r})$ introduced by the WS$_2$ layer due to the $4\%$ lattice mismatch. Thus, the continuum Hamiltonian for hole-doped TMD heterobilayers 
takes the general form
\begin{eqnarray}\label{eq_cm}
\mathcal{H} &=&\sum_i \left( - \frac{\nabla_i^2}{2m}+V(\bm r_i) \right)+  \frac{1}{2} \sum_{j \neq i} \frac{k_e \mathrm{e}^{2}}{ \epsilon\left|\mathbf{r}_i-\mathbf{r}_{j}\right|} ,
\label{eq_v}
\end{eqnarray}
where $m$ is the effective hole mass (about $0.5 m_e$ for WSe$_2$ \cite{regan2020mott,fallahazad2016shubnikov,rasmussen2015computational}), and $k_e = 1/(4\pi \epsilon_0)$ is Coulomb constant. Here $m_e$ is the electron mass and $\epsilon_0$ is the vacuum permittivity. The dielectric constant of the electrostatic environment $\epsilon$ 
is estimated to be $\epsilon \sim 6$ when  hexagonal boron nitride is used as the substrate\cite{stier2016probing}.

Since the moir\'e potential $V(\bm r)$ is smooth and has the triangular lattice symmetry, it can be parameterized using the three lowest Fourier components \cite{wu2018hubbard}: 
\begin{eqnarray}
V(\boldsymbol{r})=-2V_{0} \sum_{l=1}^{3} \cos \left(\boldsymbol{G}_{l} \cdot \boldsymbol{r}+\phi\right), \label{potential}
\end{eqnarray}
where $ \bm G_l=\frac{4\pi}{\sqrt{3} L_m}(\cos\frac{2\pi l }{3},\sin\frac{2\pi l}{3})$ are the reciprocal vectors of the moir\'e superlattice.  The moir\'e period $L_m$ increases as the twist angle $\theta$ decreases, reaching $8.2$\,nm for WSe$_2$/WS$_2$ at $\theta=0$ (aligned). 
The values of $V_0, \phi$ for various TMD heterobilayers have been obtained from our all-electron DFT calculation at charge neutrality \cite{zhang2019moir}.
For WSe$_2$/WS$_2$, $V_0 = 15$meV and $\phi \approx \pi/4$.



The continuum Hamiltonian defined by (\ref{eq_cm}) and (\ref{potential}) is the starting point of our study.
It is instructive to recast $\mathcal{H}$ in dimensionless form:
\begin{eqnarray}
{\cal H}/E_0 = \sum_i \left( - \frac{\tilde{\nabla}_i^2}{2}-  v_0 (\tilde{\bf r})  \right)+  \frac{1}{2} \sum_{j \neq i} \frac{Z}{\left|   \tilde{\mathbf{r}}_i-\tilde{\mathbf{r}}_j \right|},
\end{eqnarray}
where we define $\tilde{{\bf r}} \equiv {\bf r}/L_m$ and $E_0 \equiv \hbar^2 / (mL^2)$. The dimensionless parameter $v_0 \equiv V_0 / E_0$ characterizes the moir\'e potential strength relative to the kinetic energy, while the ratio of moir\'e period and the effective Bohr radius  $a_B =4\pi \epsilon_0 \epsilon \hbar^2 / (m e^2)$ defines the dimensionless Coulomb interaction parameter $Z \equiv L_m / a_B$.
%
The ground state thus depends on three dimensionless parameters: $v_0$, $Z$, and the filling factor $n= \frac{\sqrt{3}}{2} \rho  L_m^2$ where $\rho$ is the hole density. This makes TMD heterobilayers a highly-tunable two-dimensional hole system. Varying the moir\'e period with the twist angle and the hole density via gating can span the full range between weak and strong interaction and between weak and strong potential, thus promising a plethora of electronic states to be discovered.


We now apply the Kohn--Sham density functional theory\cite{kohn1965self} to the 2D periodic electron system defined by the continuum Hamiltonian (\ref{eq_v}). The essence of the DFT method is to find the ground state total energy $E$ and spin densities $n_\uparrow(\mathbf{r})$, $n_\downarrow(\mathbf{r})$ of an interacting system by solving an auxiliary one-body Schr\"odinger equation:
\begin{widetext}
\begin{eqnarray}
\label{equ:K-S}
 \left(-\frac{\nabla^2}{2m}+ V(\mathbf{r})+ U([n],\mathbf{r})+
\frac{\delta E_{xc}([n_\uparrow,n_\downarrow])}{\delta n_\sigma(\mathbf{r})} \right) \psi_{i\sigma}(\mathbf{r})
= \epsilon_{i\sigma} \psi_{i\sigma}(\mathbf{r})
\quad \textrm{and}   \quad
 n_\sigma (\mathbf{r})= \sum_i \theta(\mu - \epsilon_{i\sigma}) |\psi_{i\sigma} (\mathbf{r})|^2,
\end{eqnarray}
\end{widetext}
where $i\sigma$ denotes a complete set of single-particle orbitals (Kohn-Sham orbitals). Here, in addition to the external potential, the effective potential includes the Hartree potential $U$ and the spin-dependent potential $\frac{\delta E_\text{xc}([n_\uparrow,n_\downarrow])}{\delta n_\sigma(\mathbf{r})}$.  The former is associated with classical electrostatic energy of the inhomogeneous charge density
\begin{eqnarray}
U([n], \mathbf{r}) = \frac{e^2}{\epsilon} \int {\mathrm{d}}\mathbf{r}' \frac{n(\mathbf{r}')}{|\mathbf{r}-\mathbf{r}'|}, \; n(\mathbf{r}) = n_\uparrow(\mathbf{r}) + n_\downarrow(\mathbf{r}).
\end{eqnarray}
The latter depends on the exchange-correlation energy $E_\text{xc}$, which is a functional of spin-dependent  density $n_\sigma(\mathbf{r})$.
Since the effective potential itself relies on the spin density of occupied Kohn--Sham orbitals, we should solve the above equation iteratively  until reaching the self-consistency.

DFT is exact in principle, but in practice it requires an approximation to the exchange-correlation energy functional. In this work, we use the local spin density functional (LSD) approximation\cite{kohn1965self,von1972local,rajagopal1973inhomogeneous}
\begin{eqnarray}
E_\text{xc}([n_\uparrow,n_\downarrow]) = \int d \mathbf{r} \; n(\mathbf{r}) \epsilon_\text{xc}(n_\uparrow(\mathbf{r}), n_\downarrow(\mathbf{r})).
\end{eqnarray}
$\epsilon_\text{xc}(n_\uparrow, n_\downarrow)$ is the exchange-correlation energy per particle in a {\it two-dimensional} electron gas
with uniform spin densities $n_\uparrow, n_\downarrow$. In the following we use the function $\epsilon_{xc}$ obtained by Attaccalite {\it et al} from quantum Monte Carlo simulations \cite{attaccalite2002correlation}.
By construction,
the LSD approximation\cite{vosko1980accurate,perdew1981self,cole1982calculated,perdew1992accurate} is exact for homogeneous systems. In practice, DFT with the LSD approximation has proven to be accurate for a wide variety of inhomogeneous systems: atoms, molecules and natural solids including metals and 
itinerant ferromagnets \cite{moruzzi1978computed,sticht1989non}. 


\begin{figure}[t]
\includegraphics[width=1.0\linewidth]{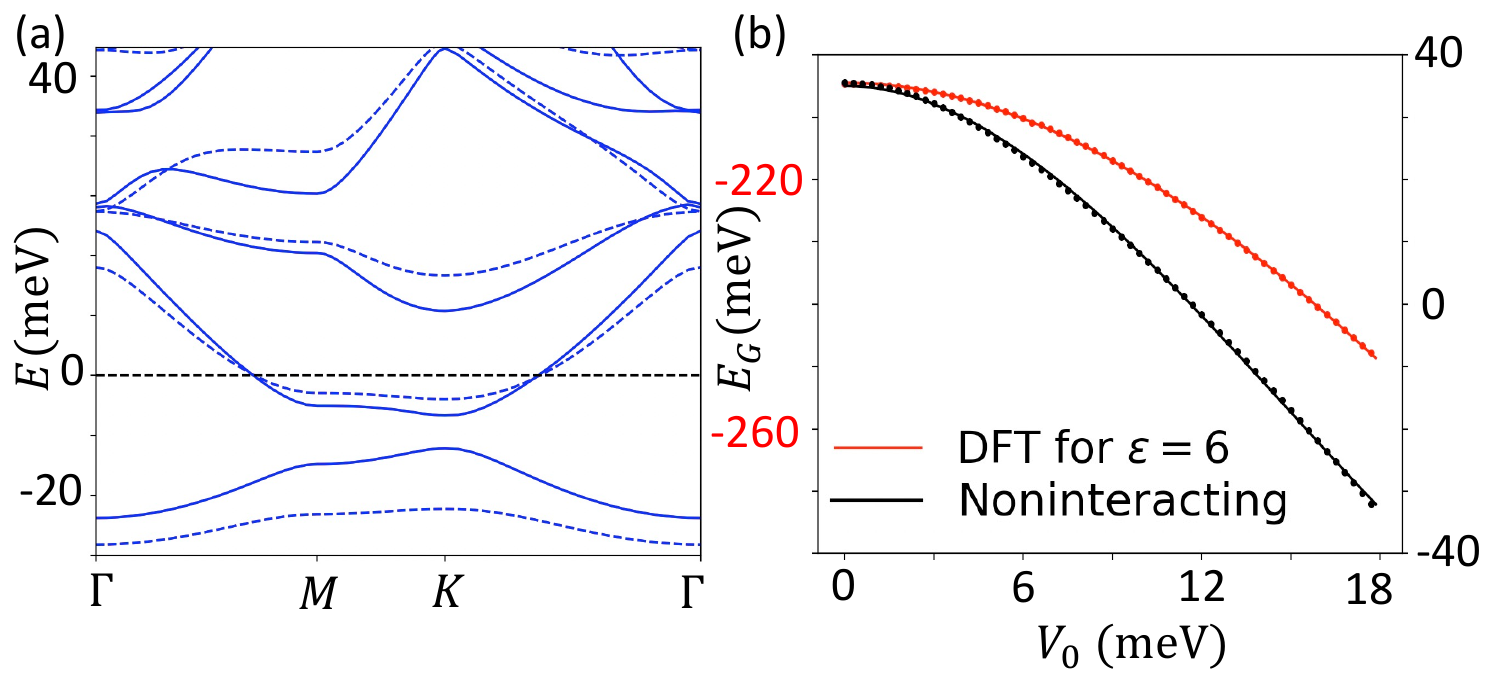}
\caption{
(a) Band structure the metallic ground state at the filling $n=3$ for $L_m=8.2$\,nm, $\phi=45^{\circ}$, obtained from DFT calculation at $\epsilon=6$ (solid line) and non-interacting continuum model (dashed line).
(b) The ground-state energy per unit cell as a function of potential strength, fitted with a fourth-order polynomial.
}
\label{fig1}
\end{figure}

\begin{figure}[t]
\includegraphics[width=1.0\linewidth]{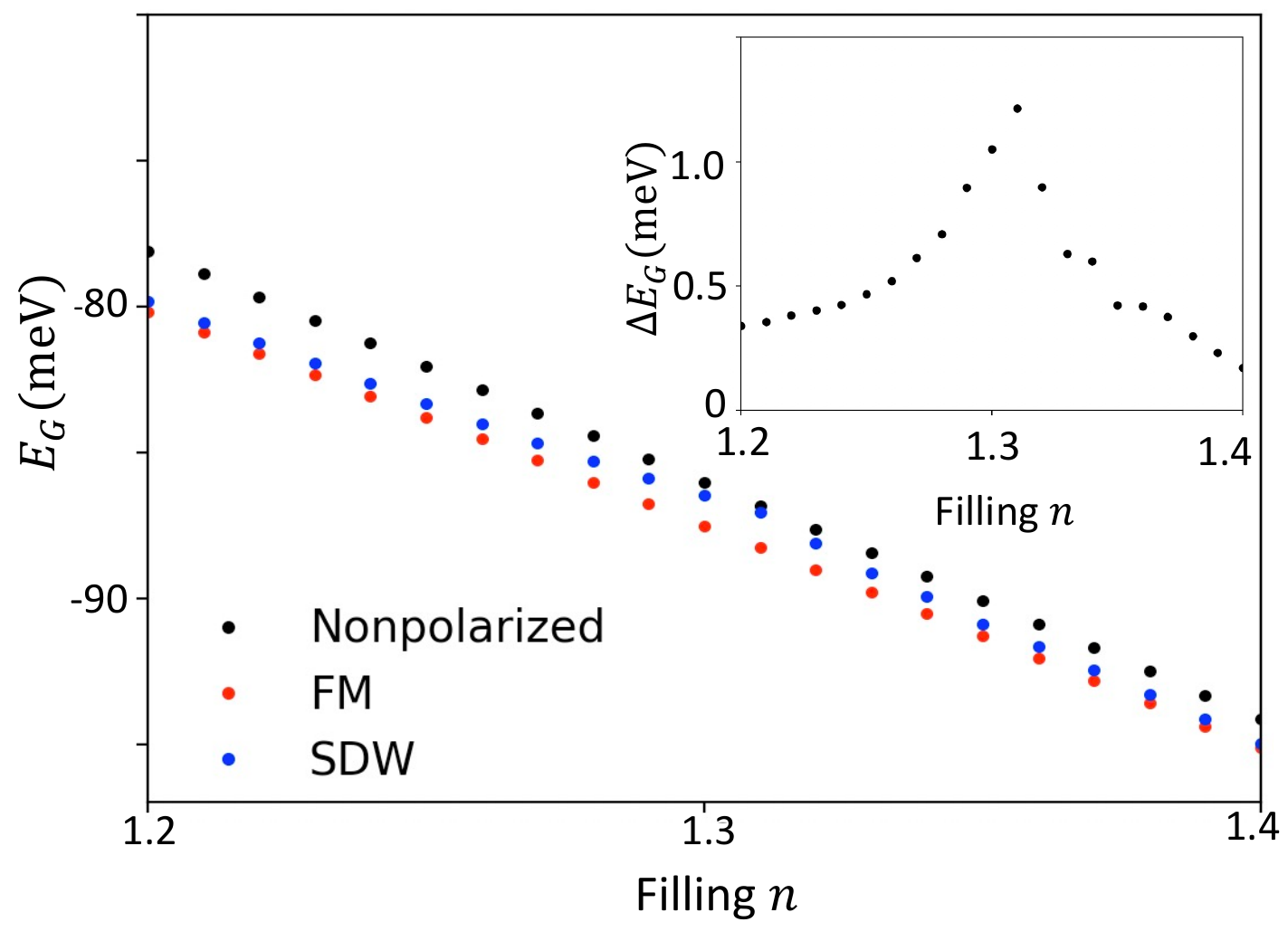}
\caption{
The ground-state energy per unit cell of fully-polarized state, spin-density-wave state, and non-spin-polarized state as a function of filling, calculated for $L_m=8.2$nm, $V_0=15\,\mathrm{meV}$, $\phi=45^\circ$ and $\epsilon=6$. The inset shows the energy difference between SDW state and FM state: $\Delta E_G=E_G(\text{SDW})-E_G(\text{FM})$.}
\label{fig2}
\end{figure}

We perform DFT calculations under LSD approximation on the continuum Hamiltonian (\ref{eq_cm},\ref{potential}) for TMD heterobilayer WSe$_2$/WS$_2$, at various charge densities and moir\'e periods $L_m$.
When the charge density is relatively high,
we find that Coulomb interaction renormalizes the metallic states at generic fillings without making qualitative changes. As an example, Figure~\ref{fig1}(a) shows the DFT band structure for aligned WSe$_2$/WS$_2$ at the filling of $n=3$ holes per moir\'e unit cell, corresponding to a hole density $\rho = 5.15 \times 10^{12}\,\mathrm{cm}^{-2}$.
We further calculate the ground-state energy $E_G$ as a function of the moir\'e potential strength $V_0$, finding an excellent fit with a fourth-order polynomial, see Fig.~\ref{fig2}(a).
This dependence,  as expected from a perturbation expansion in $V(\mathbf{r})$, proves that the metallic ground state at $V_0=15$\ meV (the parameter for WSe$_2$/WS$_2$) is adiabatically connected to the homogeneous hole gas without the moir\'e potential.

We also compare the energies and charge distributions of the metallic state at $n=3$ with and without Coulomb interaction.
Compared to the non-interacting case, the charge distribution at $\epsilon=6$ has less spatial variation, as expected from the Coulomb energy cost associated with charge inhomogeneity. 
These results on renormalized metal serves as a benchmark of our DFT method.


\begin{figure}[t]
\includegraphics[width=1.0\linewidth]{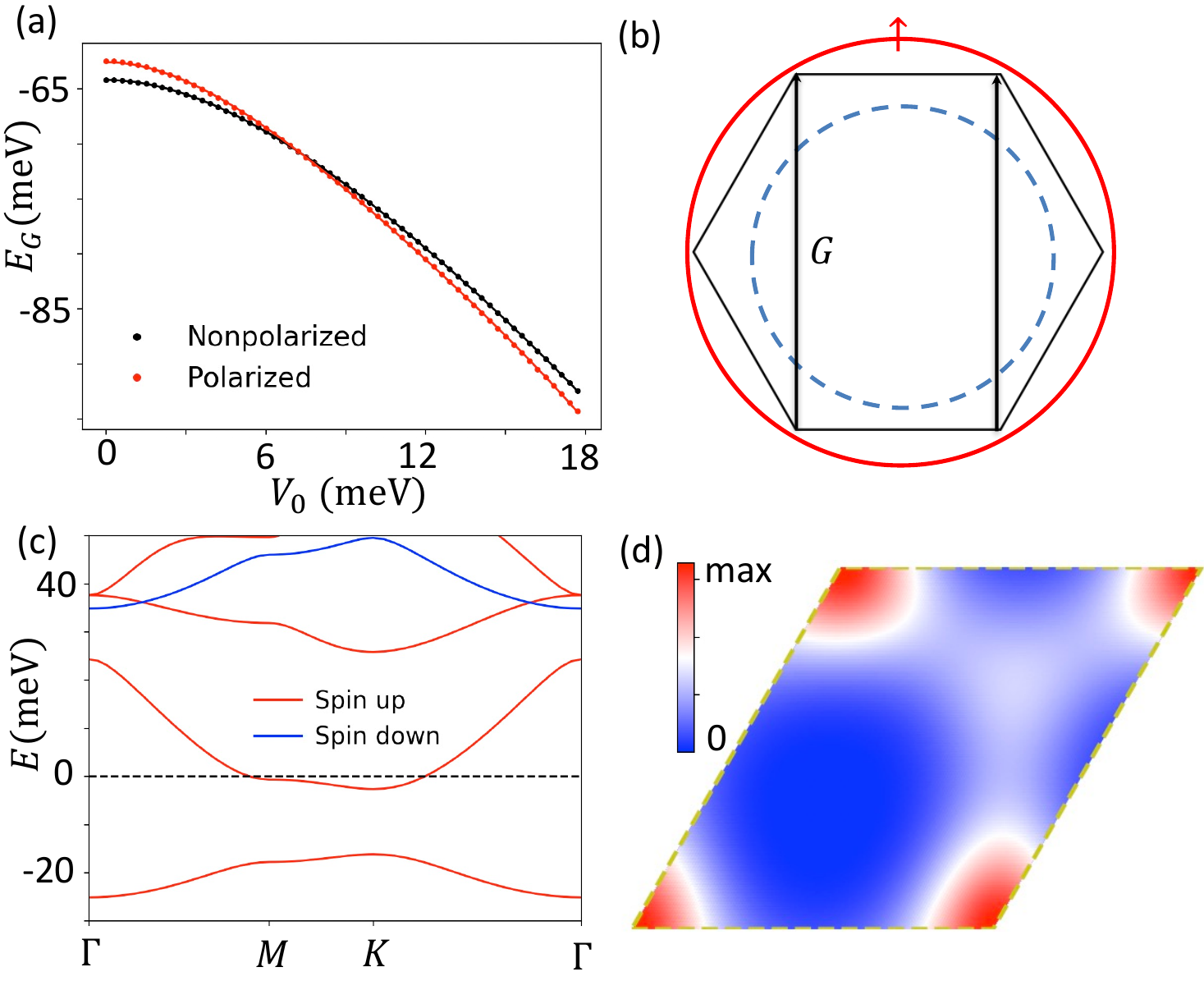}
\caption{Itinerant ferromagnet at the filling $n=1.3$ for $L_m=8.2$\,nm and $\phi=45^{\circ}$.
(a) The ground-state energy per unit cell as a function of the moir\'e potential strength $V_0$;
(b) The Fermi surfaces of the non-polarized (blue) and fully-polarized uniform electron gas (red) at the density of $n=1.3$ electrons per unit cell, in comparison with the Brillouin zone (hexagon).
(c) DFT band structure and (d) Real-space charge distribution of the fully-polarized state.}
\label{fig3}
\end{figure}

We now turn to the strongly correlated regime of TMD heterobilayers at large moir\'e wavelength and low density. For aligned WSe$_2$/WS$_2$ with $L_m=8.2$\,nm, our spin-dependent DFT calculations reveal that the ground state at fillings between $n=1.2$ and $1.4$ is a  fully spin-polarized metal.
At a given hole density $\rho$, our calculation starts from random initial spin densities $n_\uparrow(\mathbf{r})$ and $n_\downarrow(\mathbf{r})$ with average values $ \bar{n}_\uparrow = \bar{n}_\downarrow =\rho/2$. The iteration of the Kohn--Sham equation \eqref{equ:K-S}  always converges to a fully spin-polarized ground state. 

For comparison, when we enforce $n_\uparrow(\mathbf{r})=n_\downarrow(\mathbf{r})$ everywhere in solving the Kohn--Sham equation, a non-spin-polarized solution is obtained, but its energy is higher than the fully polarized state. 
We further investigate the possibility of translational symmetry breaking by performing spin-dependent DFT calculations with a $2\times 1$ enlarged unit cell. 
Starting from random initial spin densities, we find that  the majority of iterations converge to the fully-polarized metal described above, while in some cases  a metallic state with spin stripe order is obtained.

Fig.2 shows the filling-dependent ground-state energies of the fully-polarized state, the spin-density wave state, and the (enforced) non-magnetic state.
Importantly, within the filling range $1.2<n<1.4$, the fully-polarized state is always found to have the lowest energy, while the non-magnetic state has the highest energy.
Their energy difference, which measures the energy gain from itinerant ferromagnetism, decreases monotonically as the hole density increases. On the other hand, at fillings close to $1.2$ or $1.4$, the fully-polarized state and the spin-density-wave state become extremely close in energy, indicating a competition between distinct magnetic orders. The ferromagnetic ground state is the most stable around the filling $n=1.3$, where its energy per unit cell is lower than the other states by $>1$meV.


To gain insight into the origin of itinerant ferromagnetism, we calculate and compare
the ground-state energies of fully-polarized ($\zeta=1$) and non-polarized ($\zeta=0$) states  at the filling $n=1.3$, as a function of the potential strength $V_0$.
Here $\xi \equiv (N_\uparrow - N_\downarrow)/(N_\uparrow + N_\downarrow)$ denotes the spin polarization.
As shown in Figure~\ref{fig3}(a), $E^{\zeta=1,0}(V_0)$ fit nicely with fourth-order polynomials, which proves the adiabatic continuity to the uniform hole gas at $V_0=0$, with full and zero spin polarization respectively.

At $V_0=0$, the non-polarized uniform hole gas has a lower energy. However, due to the large effective mass and low density (the corresponding interaction parameter is $r_s=5.35$), its spin susceptibility is strongly enhanced by Coulomb interaction, making its energy difference with the fully-polarized state relatively small. 
At the hole density corresponding to the filling $n=1.3$, the Fermi surface in the fully-polarized case lies just slightly outside the first Brillouin zone (a hexagon), and in the non-polarized case falls entirely within it, i.e., $2k_F^{\xi=0}<G<2k_F^{\xi=1}$ as shown in Fig.~\ref{fig2}(d). Since the density wave susceptibility of a uniform hole gas decreases rapidly at wavevectors greater than $2k_F$, the fully-polarized state exhibits a larger density response to the moir\'e potential, hence lowers its energy faster with increasing potential strength, and eventually becomes the true ground state above a critical $V_0$.


%

\begin{figure}
\includegraphics[width=0.8\linewidth]{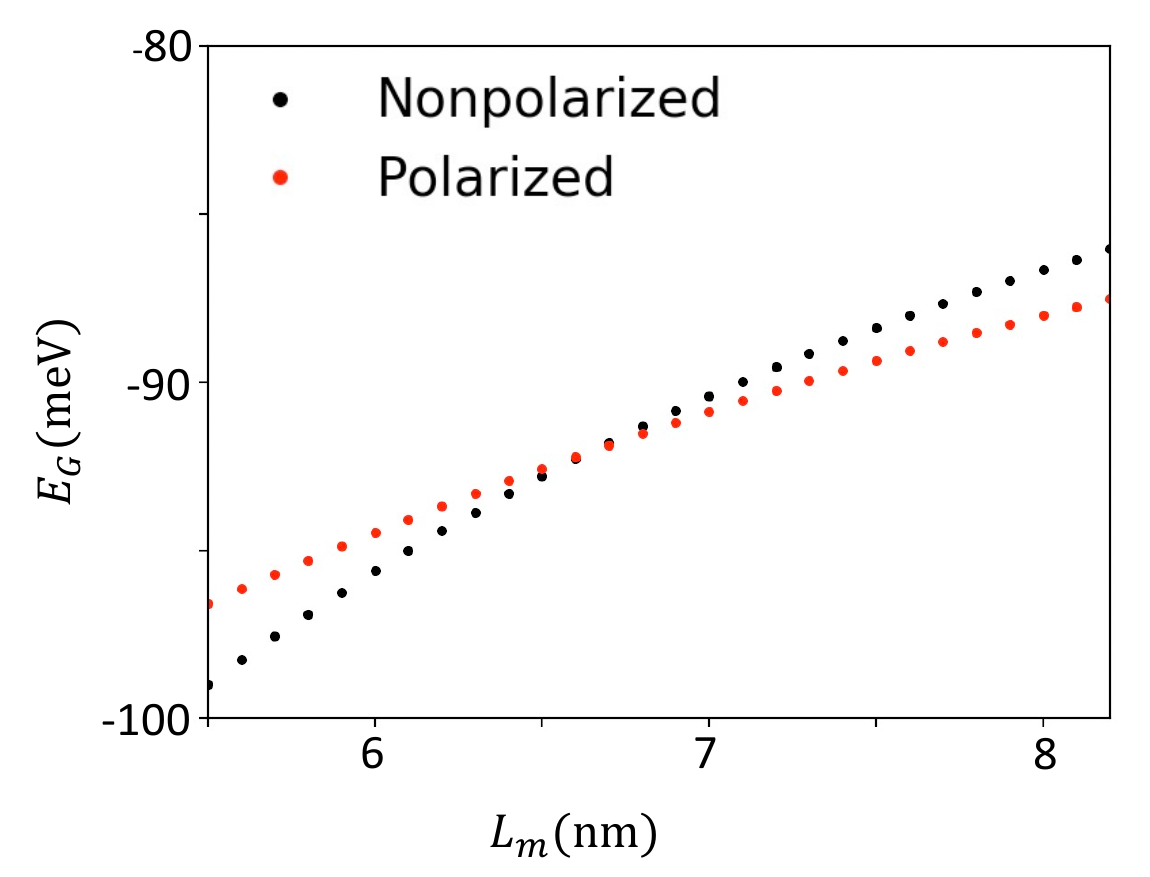}
\caption{
The ground-state energy per unit cell of the fully-polarized and non-polarized state as a function of the moir\'e period, at the filling $n=1.3$, $V_0=15\,\mathrm{meV}$, and $\phi=45^\circ$}
\label{fig4}
\end{figure}

The electronic structure and real-space charge distribution of the ferromagnetic (FM) metal at $n=1.3$ are shown in Fig.~\ref{fig1}(c) and (d) respectively.  The lowest moir\'e band, completely filled by spin-polarized holes, give rise to one $s=1/2$ local moment in the $AA$ region of each moir\'e unit cell. The remaining $0.3$ holes per unit cell partially occupy the second moir\'e band, and are distributed primarily in the $AB$ region, consistent with the strong-coupling picture of  $n-1$ doped holes being transferred to $AB$ orbitals to avoid double occupancy of $AA$ orbitals \cite{zhang2019moir}. It is worth emphasizing that in the ferromagnetic metal we found at $n>1$, the (pseudo-)spins of {\it all} holes are aligned, resulting in $n$ pseudospin-$1/2$  magnetic moment per moir\'e unit cell, whereas the density of mobile charge carriers is $n-1$.  This behavior cannot be captured by any single-band Hubbard model, which can accommodate at most  $2-n$ spins per unit cell at $n>1$.

We further show that itinerant ferromagnetism in TMD heterobilayers is tunable by increasing the twist angle $\theta$ to reduce the moir\'e period $L_m$. A smaller $L_m$ leads to a larger bandwidth and hence weakens electron correlation.  Figure~\ref{fig4} shows the energies of the FM state and the non-magnetic state as a function of $L_m$. A quantum phase transition from the FM state to the non-magnetic state occurs at $L_m \approx 6.5\,\mathrm{nm}$, or $\theta \approx 1.75^\circ$.

We now compare our findings with a recent optical experiment on hole-doped WSe$_2$/WS$_2$, which observed Curie-Weiss behavior in the temperature dependence of the exciton Zeeman splitting. Remarkably,  above the filling $n=1.2$,  the Weiss constant changes sign from negative to positive, consistent with a ferromagnetic metal we found between $n=1.2$ and $1.4$. To establish ferromagnetism in WSe$_2$/WS$_2$, it is desirable to measure the magnetization as a function of the filling by SQUID. The reduced density $n-1$ of mobile holes in the FM metal can be measured in Hall and quantum oscillation experiments. The complete removal of spin degeneracy can be directly tested from the Landau level fan diagram.





At nonzero temperature, thermal fluctuations destroy long-range FM order in two dimensional systems with the continuous spin-rotational symmetry. For TMD heterobilayers, however, magnetic dipole-dipole interaction or higher-order valley-contrasting energy dispersion such as $(k_x^3- 3 k_x k_y^2) s_z$, is expected to result in uniaxial magnetic anisotropy and thus stabilize the itinerant ferromagnetism at low temperature. If the $z$ direction is the easy axis, the FM metal has a spontaneous out-of-plane magnetization, which can be detected by anomalous Hall effect or magnetic circular dichroism. Alternatively, an easy-plane anisotropy will lead to an inter-valley coherent state, which is expected to enable superfluid-like spin transport \cite{bi2019excitonic,jin2018imaging}. Thus, TMD moir\'e superlattices provide a new and  highly tunable platform for the study of fundamental mechanism of itinerant magnetism in two dimensions, and may enable the design of electrically controllable spintronics devices \cite{huang2018electrical,jiang2018controlling,manchon2019current}.


To summarize, our DFT calculations reveal itinerant ferromagnetism stabilized by the moir\'e potential in a two-dimensional Coulomb system, the TMD-based moir\'e superlattices. This finding, which goes beyond the scope of single-band Hubbard model, demonstrates the predictive power of our moir\'e-DFT method targeting directly low-energy electrons in a coarse-grained moir\'e potential. This method will discover a variety of broken-symmetry states in TMD heterobilayers at other fillings, thus guiding future experiments.
Its capability  will be further  extended by improvements in the exchange-correlation functional to 
handle Dirac electrons in graphene-based moir\'e systems. We hope that moir\'e-DFT becomes
a useful tool for the investigation of many-electron physics in all moir\'e materials.

\section*{Acknowledgment}
We thank Kin Fai Mak, Jie Shan and  Feng Wang for numerous discussions on  experiments, and Allan MacDonald for valuable theoretical discussions.

\bibliography{dft}

\begin{thebibliography}{46}
\expandafter\ifx\csname natexlab\endcsname\relax\def\natexlab#1{#1}\fi
\expandafter\ifx\csname bibnamefont\endcsname\relax
  \def\bibnamefont#1{#1}\fi
\expandafter\ifx\csname bibfnamefont\endcsname\relax
  \def\bibfnamefont#1{#1}\fi
\expandafter\ifx\csname citenamefont\endcsname\relax
  \def\citenamefont#1{#1}\fi
\expandafter\ifx\csname url\endcsname\relax
  \def\url#1{\texttt{#1}}\fi
\expandafter\ifx\csname urlprefix\endcsname\relax\def\urlprefix{URL }\fi
\providecommand{\bibinfo}[2]{#2}
\providecommand{\eprint}[2][]{\url{#2}}

\bibitem[{\citenamefont{Cao et~al.}(2018{\natexlab{a}})\citenamefont{Cao,
  Fatemi, Demir, Fang, Tomarken, Luo, Sanchez-Yamagishi, Watanabe, Taniguchi,
  Kaxiras et~al.}}]{cao2018correlated}
\bibinfo{author}{\bibfnamefont{Y.}~\bibnamefont{Cao}},
  \bibinfo{author}{\bibfnamefont{V.}~\bibnamefont{Fatemi}},
  \bibinfo{author}{\bibfnamefont{A.}~\bibnamefont{Demir}},
  \bibinfo{author}{\bibfnamefont{S.}~\bibnamefont{Fang}},
  \bibinfo{author}{\bibfnamefont{S.~L.} \bibnamefont{Tomarken}},
  \bibinfo{author}{\bibfnamefont{J.~Y.} \bibnamefont{Luo}},
  \bibinfo{author}{\bibfnamefont{J.~D.} \bibnamefont{Sanchez-Yamagishi}},
  \bibinfo{author}{\bibfnamefont{K.}~\bibnamefont{Watanabe}},
  \bibinfo{author}{\bibfnamefont{T.}~\bibnamefont{Taniguchi}},
  \bibinfo{author}{\bibfnamefont{E.}~\bibnamefont{Kaxiras}},
  \bibnamefont{et~al.}, \bibinfo{journal}{Nature}
  \textbf{\bibinfo{volume}{556}}, \bibinfo{pages}{80}
  (\bibinfo{year}{2018}{\natexlab{a}}).

\bibitem[{\citenamefont{Cao et~al.}(2018{\natexlab{b}})\citenamefont{Cao,
  Fatemi, Fang, Watanabe, Taniguchi, Kaxiras, and
  Jarillo-Herrero}}]{cao2018unconventional}
\bibinfo{author}{\bibfnamefont{Y.}~\bibnamefont{Cao}},
  \bibinfo{author}{\bibfnamefont{V.}~\bibnamefont{Fatemi}},
  \bibinfo{author}{\bibfnamefont{S.}~\bibnamefont{Fang}},
  \bibinfo{author}{\bibfnamefont{K.}~\bibnamefont{Watanabe}},
  \bibinfo{author}{\bibfnamefont{T.}~\bibnamefont{Taniguchi}},
  \bibinfo{author}{\bibfnamefont{E.}~\bibnamefont{Kaxiras}}, \bibnamefont{and}
  \bibinfo{author}{\bibfnamefont{P.}~\bibnamefont{Jarillo-Herrero}},
  \bibinfo{journal}{Nature} \textbf{\bibinfo{volume}{556}}, \bibinfo{pages}{43}
  (\bibinfo{year}{2018}{\natexlab{b}}).

\bibitem[{\citenamefont{Lu et~al.}(2019)\citenamefont{Lu, Stepanov, Yang, Xie,
  Aamir, Das, Urgell, Watanabe, Taniguchi, Zhang
  et~al.}}]{lu2019superconductors}
\bibinfo{author}{\bibfnamefont{X.}~\bibnamefont{Lu}},
  \bibinfo{author}{\bibfnamefont{P.}~\bibnamefont{Stepanov}},
  \bibinfo{author}{\bibfnamefont{W.}~\bibnamefont{Yang}},
  \bibinfo{author}{\bibfnamefont{M.}~\bibnamefont{Xie}},
  \bibinfo{author}{\bibfnamefont{M.~A.} \bibnamefont{Aamir}},
  \bibinfo{author}{\bibfnamefont{I.}~\bibnamefont{Das}},
  \bibinfo{author}{\bibfnamefont{C.}~\bibnamefont{Urgell}},
  \bibinfo{author}{\bibfnamefont{K.}~\bibnamefont{Watanabe}},
  \bibinfo{author}{\bibfnamefont{T.}~\bibnamefont{Taniguchi}},
  \bibinfo{author}{\bibfnamefont{G.}~\bibnamefont{Zhang}},
  \bibnamefont{et~al.}, \bibinfo{journal}{Nature}
  \textbf{\bibinfo{volume}{574}}, \bibinfo{pages}{653} (\bibinfo{year}{2019}).

\bibitem[{\citenamefont{Kerelsky et~al.}(2019)\citenamefont{Kerelsky, McGilly,
  Kennes, Xian, Yankowitz, Chen, Watanabe, Taniguchi, Hone, Dean
  et~al.}}]{kerelsky2019maximized}
\bibinfo{author}{\bibfnamefont{A.}~\bibnamefont{Kerelsky}},
  \bibinfo{author}{\bibfnamefont{L.~J.} \bibnamefont{McGilly}},
  \bibinfo{author}{\bibfnamefont{D.~M.} \bibnamefont{Kennes}},
  \bibinfo{author}{\bibfnamefont{L.}~\bibnamefont{Xian}},
  \bibinfo{author}{\bibfnamefont{M.}~\bibnamefont{Yankowitz}},
  \bibinfo{author}{\bibfnamefont{S.}~\bibnamefont{Chen}},
  \bibinfo{author}{\bibfnamefont{K.}~\bibnamefont{Watanabe}},
  \bibinfo{author}{\bibfnamefont{T.}~\bibnamefont{Taniguchi}},
  \bibinfo{author}{\bibfnamefont{J.}~\bibnamefont{Hone}},
  \bibinfo{author}{\bibfnamefont{C.}~\bibnamefont{Dean}}, \bibnamefont{et~al.},
  \bibinfo{journal}{Nature} \textbf{\bibinfo{volume}{572}}, \bibinfo{pages}{95}
  (\bibinfo{year}{2019}).

\bibitem[{\citenamefont{Jiang et~al.}(2019)\citenamefont{Jiang, Lai, Watanabe,
  Taniguchi, Haule, Mao, and Andrei}}]{jiang2019charge}
\bibinfo{author}{\bibfnamefont{Y.}~\bibnamefont{Jiang}},
  \bibinfo{author}{\bibfnamefont{X.}~\bibnamefont{Lai}},
  \bibinfo{author}{\bibfnamefont{K.}~\bibnamefont{Watanabe}},
  \bibinfo{author}{\bibfnamefont{T.}~\bibnamefont{Taniguchi}},
  \bibinfo{author}{\bibfnamefont{K.}~\bibnamefont{Haule}},
  \bibinfo{author}{\bibfnamefont{J.}~\bibnamefont{Mao}}, \bibnamefont{and}
  \bibinfo{author}{\bibfnamefont{E.~Y.} \bibnamefont{Andrei}},
  \bibinfo{journal}{Nature} \textbf{\bibinfo{volume}{573}}, \bibinfo{pages}{91}
  (\bibinfo{year}{2019}).

\bibitem[{\citenamefont{Xie et~al.}(2019)\citenamefont{Xie, Lian, J{\"a}ck,
  Liu, Chiu, Watanabe, Taniguchi, Bernevig, and
  Yazdani}}]{xie2019spectroscopic}
\bibinfo{author}{\bibfnamefont{Y.}~\bibnamefont{Xie}},
  \bibinfo{author}{\bibfnamefont{B.}~\bibnamefont{Lian}},
  \bibinfo{author}{\bibfnamefont{B.}~\bibnamefont{J{\"a}ck}},
  \bibinfo{author}{\bibfnamefont{X.}~\bibnamefont{Liu}},
  \bibinfo{author}{\bibfnamefont{C.-L.} \bibnamefont{Chiu}},
  \bibinfo{author}{\bibfnamefont{K.}~\bibnamefont{Watanabe}},
  \bibinfo{author}{\bibfnamefont{T.}~\bibnamefont{Taniguchi}},
  \bibinfo{author}{\bibfnamefont{B.~A.} \bibnamefont{Bernevig}},
  \bibnamefont{and} \bibinfo{author}{\bibfnamefont{A.}~\bibnamefont{Yazdani}},
  \bibinfo{journal}{Nature} \textbf{\bibinfo{volume}{572}},
  \bibinfo{pages}{101} (\bibinfo{year}{2019}).

\bibitem[{\citenamefont{Choi et~al.}(2019)\citenamefont{Choi, Kemmer, Peng,
  Thomson, Arora, Polski, Zhang, Ren, Alicea, Refael
  et~al.}}]{choi2019electronic}
\bibinfo{author}{\bibfnamefont{Y.}~\bibnamefont{Choi}},
  \bibinfo{author}{\bibfnamefont{J.}~\bibnamefont{Kemmer}},
  \bibinfo{author}{\bibfnamefont{Y.}~\bibnamefont{Peng}},
  \bibinfo{author}{\bibfnamefont{A.}~\bibnamefont{Thomson}},
  \bibinfo{author}{\bibfnamefont{H.}~\bibnamefont{Arora}},
  \bibinfo{author}{\bibfnamefont{R.}~\bibnamefont{Polski}},
  \bibinfo{author}{\bibfnamefont{Y.}~\bibnamefont{Zhang}},
  \bibinfo{author}{\bibfnamefont{H.}~\bibnamefont{Ren}},
  \bibinfo{author}{\bibfnamefont{J.}~\bibnamefont{Alicea}},
  \bibinfo{author}{\bibfnamefont{G.}~\bibnamefont{Refael}},
  \bibnamefont{et~al.}, \bibinfo{journal}{Nature Physics}
  \textbf{\bibinfo{volume}{15}}, \bibinfo{pages}{1174} (\bibinfo{year}{2019}).

\bibitem[{\citenamefont{Yankowitz et~al.}(2019)\citenamefont{Yankowitz, Chen,
  Polshyn, Zhang, Watanabe, Taniguchi, Graf, Young, and
  Dean}}]{yankowitz2019tuning}
\bibinfo{author}{\bibfnamefont{M.}~\bibnamefont{Yankowitz}},
  \bibinfo{author}{\bibfnamefont{S.}~\bibnamefont{Chen}},
  \bibinfo{author}{\bibfnamefont{H.}~\bibnamefont{Polshyn}},
  \bibinfo{author}{\bibfnamefont{Y.}~\bibnamefont{Zhang}},
  \bibinfo{author}{\bibfnamefont{K.}~\bibnamefont{Watanabe}},
  \bibinfo{author}{\bibfnamefont{T.}~\bibnamefont{Taniguchi}},
  \bibinfo{author}{\bibfnamefont{D.}~\bibnamefont{Graf}},
  \bibinfo{author}{\bibfnamefont{A.~F.} \bibnamefont{Young}}, \bibnamefont{and}
  \bibinfo{author}{\bibfnamefont{C.~R.} \bibnamefont{Dean}},
  \bibinfo{journal}{Science} \textbf{\bibinfo{volume}{363}},
  \bibinfo{pages}{1059} (\bibinfo{year}{2019}).

\bibitem[{\citenamefont{Codecido et~al.}(2019)\citenamefont{Codecido, Wang,
  Koester, Che, Tian, Lv, Tran, Watanabe, Taniguchi, Zhang
  et~al.}}]{codecido2019correlated}
\bibinfo{author}{\bibfnamefont{E.}~\bibnamefont{Codecido}},
  \bibinfo{author}{\bibfnamefont{Q.}~\bibnamefont{Wang}},
  \bibinfo{author}{\bibfnamefont{R.}~\bibnamefont{Koester}},
  \bibinfo{author}{\bibfnamefont{S.}~\bibnamefont{Che}},
  \bibinfo{author}{\bibfnamefont{H.}~\bibnamefont{Tian}},
  \bibinfo{author}{\bibfnamefont{R.}~\bibnamefont{Lv}},
  \bibinfo{author}{\bibfnamefont{S.}~\bibnamefont{Tran}},
  \bibinfo{author}{\bibfnamefont{K.}~\bibnamefont{Watanabe}},
  \bibinfo{author}{\bibfnamefont{T.}~\bibnamefont{Taniguchi}},
  \bibinfo{author}{\bibfnamefont{F.}~\bibnamefont{Zhang}},
  \bibnamefont{et~al.}, \bibinfo{journal}{Science Advances}
  \textbf{\bibinfo{volume}{5}}, \bibinfo{pages}{eaaw9770}
  (\bibinfo{year}{2019}).

\bibitem[{\citenamefont{Sharpe et~al.}(2019)\citenamefont{Sharpe, Fox, Barnard,
  Finney, Watanabe, Taniguchi, Kastner, and
  Goldhaber-Gordon}}]{sharpe2019emergent}
\bibinfo{author}{\bibfnamefont{A.~L.} \bibnamefont{Sharpe}},
  \bibinfo{author}{\bibfnamefont{E.~J.} \bibnamefont{Fox}},
  \bibinfo{author}{\bibfnamefont{A.~W.} \bibnamefont{Barnard}},
  \bibinfo{author}{\bibfnamefont{J.}~\bibnamefont{Finney}},
  \bibinfo{author}{\bibfnamefont{K.}~\bibnamefont{Watanabe}},
  \bibinfo{author}{\bibfnamefont{T.}~\bibnamefont{Taniguchi}},
  \bibinfo{author}{\bibfnamefont{M.}~\bibnamefont{Kastner}}, \bibnamefont{and}
  \bibinfo{author}{\bibfnamefont{D.}~\bibnamefont{Goldhaber-Gordon}},
  \bibinfo{journal}{Science} \textbf{\bibinfo{volume}{365}},
  \bibinfo{pages}{605} (\bibinfo{year}{2019}).

\bibitem[{\citenamefont{Tomarken et~al.}(2019)\citenamefont{Tomarken, Cao,
  Demir, Watanabe, Taniguchi, Jarillo-Herrero, and
  Ashoori}}]{tomarken2019electronic}
\bibinfo{author}{\bibfnamefont{S.}~\bibnamefont{Tomarken}},
  \bibinfo{author}{\bibfnamefont{Y.}~\bibnamefont{Cao}},
  \bibinfo{author}{\bibfnamefont{A.}~\bibnamefont{Demir}},
  \bibinfo{author}{\bibfnamefont{K.}~\bibnamefont{Watanabe}},
  \bibinfo{author}{\bibfnamefont{T.}~\bibnamefont{Taniguchi}},
  \bibinfo{author}{\bibfnamefont{P.}~\bibnamefont{Jarillo-Herrero}},
  \bibnamefont{and} \bibinfo{author}{\bibfnamefont{R.}~\bibnamefont{Ashoori}},
  \bibinfo{journal}{Physical review letters} \textbf{\bibinfo{volume}{123}},
  \bibinfo{pages}{046601} (\bibinfo{year}{2019}).

\bibitem[{\citenamefont{Zondiner et~al.}(2019)\citenamefont{Zondiner, Rozen,
  Rodan-Legrain, Cao, Queiroz, Taniguchi, Watanabe, Oreg, von Oppen, Stern
  et~al.}}]{zondiner2019cascade}
\bibinfo{author}{\bibfnamefont{U.}~\bibnamefont{Zondiner}},
  \bibinfo{author}{\bibfnamefont{A.}~\bibnamefont{Rozen}},
  \bibinfo{author}{\bibfnamefont{D.}~\bibnamefont{Rodan-Legrain}},
  \bibinfo{author}{\bibfnamefont{Y.}~\bibnamefont{Cao}},
  \bibinfo{author}{\bibfnamefont{R.}~\bibnamefont{Queiroz}},
  \bibinfo{author}{\bibfnamefont{T.}~\bibnamefont{Taniguchi}},
  \bibinfo{author}{\bibfnamefont{K.}~\bibnamefont{Watanabe}},
  \bibinfo{author}{\bibfnamefont{Y.}~\bibnamefont{Oreg}},
  \bibinfo{author}{\bibfnamefont{F.}~\bibnamefont{von Oppen}},
  \bibinfo{author}{\bibfnamefont{A.}~\bibnamefont{Stern}},
  \bibnamefont{et~al.}, \bibinfo{journal}{arXiv preprint arXiv:1912.06150}
  (\bibinfo{year}{2019}).

\bibitem[{\citenamefont{Chen et~al.}(2019{\natexlab{a}})\citenamefont{Chen,
  Jiang, Wu, Lyu, Li, Chittari, Watanabe, Taniguchi, Shi, Jung
  et~al.}}]{chen2019evidence}
\bibinfo{author}{\bibfnamefont{G.}~\bibnamefont{Chen}},
  \bibinfo{author}{\bibfnamefont{L.}~\bibnamefont{Jiang}},
  \bibinfo{author}{\bibfnamefont{S.}~\bibnamefont{Wu}},
  \bibinfo{author}{\bibfnamefont{B.}~\bibnamefont{Lyu}},
  \bibinfo{author}{\bibfnamefont{H.}~\bibnamefont{Li}},
  \bibinfo{author}{\bibfnamefont{B.~L.} \bibnamefont{Chittari}},
  \bibinfo{author}{\bibfnamefont{K.}~\bibnamefont{Watanabe}},
  \bibinfo{author}{\bibfnamefont{T.}~\bibnamefont{Taniguchi}},
  \bibinfo{author}{\bibfnamefont{Z.}~\bibnamefont{Shi}},
  \bibinfo{author}{\bibfnamefont{J.}~\bibnamefont{Jung}}, \bibnamefont{et~al.},
  \bibinfo{journal}{Nature Physics} \textbf{\bibinfo{volume}{15}},
  \bibinfo{pages}{237} (\bibinfo{year}{2019}{\natexlab{a}}).

\bibitem[{\citenamefont{Chen et~al.}(2019{\natexlab{b}})\citenamefont{Chen,
  Sharpe, Gallagher, Rosen, Fox, Jiang, Lyu, Li, Watanabe, Taniguchi
  et~al.}}]{chen2019sig}
\bibinfo{author}{\bibfnamefont{G.}~\bibnamefont{Chen}},
  \bibinfo{author}{\bibfnamefont{A.~L.} \bibnamefont{Sharpe}},
  \bibinfo{author}{\bibfnamefont{P.}~\bibnamefont{Gallagher}},
  \bibinfo{author}{\bibfnamefont{I.~T.} \bibnamefont{Rosen}},
  \bibinfo{author}{\bibfnamefont{E.~J.} \bibnamefont{Fox}},
  \bibinfo{author}{\bibfnamefont{L.}~\bibnamefont{Jiang}},
  \bibinfo{author}{\bibfnamefont{B.}~\bibnamefont{Lyu}},
  \bibinfo{author}{\bibfnamefont{H.}~\bibnamefont{Li}},
  \bibinfo{author}{\bibfnamefont{K.}~\bibnamefont{Watanabe}},
  \bibinfo{author}{\bibfnamefont{T.}~\bibnamefont{Taniguchi}},
  \bibnamefont{et~al.}, \bibinfo{journal}{Nature}
  \textbf{\bibinfo{volume}{572}}, \bibinfo{pages}{215}
  (\bibinfo{year}{2019}{\natexlab{b}}),
  \urlprefix\url{https://doi.org/10.1038/s41586-019-1393-y}.

\bibitem[{\citenamefont{Serlin et~al.}(2019)\citenamefont{Serlin, Tschirhart,
  Polshyn, Zhang, Zhu, Watanabe, Taniguchi, Balents, and
  Young}}]{serlin2019intrinsic}
\bibinfo{author}{\bibfnamefont{M.}~\bibnamefont{Serlin}},
  \bibinfo{author}{\bibfnamefont{C.}~\bibnamefont{Tschirhart}},
  \bibinfo{author}{\bibfnamefont{H.}~\bibnamefont{Polshyn}},
  \bibinfo{author}{\bibfnamefont{Y.}~\bibnamefont{Zhang}},
  \bibinfo{author}{\bibfnamefont{J.}~\bibnamefont{Zhu}},
  \bibinfo{author}{\bibfnamefont{K.}~\bibnamefont{Watanabe}},
  \bibinfo{author}{\bibfnamefont{T.}~\bibnamefont{Taniguchi}},
  \bibinfo{author}{\bibfnamefont{L.}~\bibnamefont{Balents}}, \bibnamefont{and}
  \bibinfo{author}{\bibfnamefont{A.}~\bibnamefont{Young}},
  \bibinfo{journal}{Science}  (\bibinfo{year}{2019}).

\bibitem[{\citenamefont{Liu et~al.}(2019)\citenamefont{Liu, Hao, Khalaf, Lee,
  Watanabe, Taniguchi, Vishwanath, and Kim}}]{liu2019spin}
\bibinfo{author}{\bibfnamefont{X.}~\bibnamefont{Liu}},
  \bibinfo{author}{\bibfnamefont{Z.}~\bibnamefont{Hao}},
  \bibinfo{author}{\bibfnamefont{E.}~\bibnamefont{Khalaf}},
  \bibinfo{author}{\bibfnamefont{J.~Y.} \bibnamefont{Lee}},
  \bibinfo{author}{\bibfnamefont{K.}~\bibnamefont{Watanabe}},
  \bibinfo{author}{\bibfnamefont{T.}~\bibnamefont{Taniguchi}},
  \bibinfo{author}{\bibfnamefont{A.}~\bibnamefont{Vishwanath}},
  \bibnamefont{and} \bibinfo{author}{\bibfnamefont{P.}~\bibnamefont{Kim}},
  \bibinfo{journal}{arXiv preprint arXiv:1903.08130}  (\bibinfo{year}{2019}).

\bibitem[{\citenamefont{Tang et~al.}(2020)\citenamefont{Tang, Li, Li, Xu, Liu,
  Barmak, Watanabe, Taniguchi, MacDonald, Shan et~al.}}]{tang2020simulation}
\bibinfo{author}{\bibfnamefont{Y.}~\bibnamefont{Tang}},
  \bibinfo{author}{\bibfnamefont{L.}~\bibnamefont{Li}},
  \bibinfo{author}{\bibfnamefont{T.}~\bibnamefont{Li}},
  \bibinfo{author}{\bibfnamefont{Y.}~\bibnamefont{Xu}},
  \bibinfo{author}{\bibfnamefont{S.}~\bibnamefont{Liu}},
  \bibinfo{author}{\bibfnamefont{K.}~\bibnamefont{Barmak}},
  \bibinfo{author}{\bibfnamefont{K.}~\bibnamefont{Watanabe}},
  \bibinfo{author}{\bibfnamefont{T.}~\bibnamefont{Taniguchi}},
  \bibinfo{author}{\bibfnamefont{A.~H.} \bibnamefont{MacDonald}},
  \bibinfo{author}{\bibfnamefont{J.}~\bibnamefont{Shan}}, \bibnamefont{et~al.},
  \bibinfo{journal}{Nature} \textbf{\bibinfo{volume}{579}},
  \bibinfo{pages}{353} (\bibinfo{year}{2020}).

\bibitem[{\citenamefont{Regan et~al.}(2020)\citenamefont{Regan, Wang, Jin,
  Utama, Gao, Wei, Zhao, Zhao, Zhang, Yumigeta et~al.}}]{regan2020mott}
\bibinfo{author}{\bibfnamefont{E.~C.} \bibnamefont{Regan}},
  \bibinfo{author}{\bibfnamefont{D.}~\bibnamefont{Wang}},
  \bibinfo{author}{\bibfnamefont{C.}~\bibnamefont{Jin}},
  \bibinfo{author}{\bibfnamefont{M.~I.~B.} \bibnamefont{Utama}},
  \bibinfo{author}{\bibfnamefont{B.}~\bibnamefont{Gao}},
  \bibinfo{author}{\bibfnamefont{X.}~\bibnamefont{Wei}},
  \bibinfo{author}{\bibfnamefont{S.}~\bibnamefont{Zhao}},
  \bibinfo{author}{\bibfnamefont{W.}~\bibnamefont{Zhao}},
  \bibinfo{author}{\bibfnamefont{Z.}~\bibnamefont{Zhang}},
  \bibinfo{author}{\bibfnamefont{K.}~\bibnamefont{Yumigeta}},
  \bibnamefont{et~al.}, \bibinfo{journal}{Nature}
  \textbf{\bibinfo{volume}{579}}, \bibinfo{pages}{359} (\bibinfo{year}{2020}).

\bibitem[{\citenamefont{Wang et~al.}(2019)\citenamefont{Wang, Shih, Ghiotto,
  Xian, Rhodes, Tan, Claassen, Kennes, Bai, Kim et~al.}}]{wang2019magic}
\bibinfo{author}{\bibfnamefont{L.}~\bibnamefont{Wang}},
  \bibinfo{author}{\bibfnamefont{E.-M.} \bibnamefont{Shih}},
  \bibinfo{author}{\bibfnamefont{A.}~\bibnamefont{Ghiotto}},
  \bibinfo{author}{\bibfnamefont{L.}~\bibnamefont{Xian}},
  \bibinfo{author}{\bibfnamefont{D.~A.} \bibnamefont{Rhodes}},
  \bibinfo{author}{\bibfnamefont{C.}~\bibnamefont{Tan}},
  \bibinfo{author}{\bibfnamefont{M.}~\bibnamefont{Claassen}},
  \bibinfo{author}{\bibfnamefont{D.~M.} \bibnamefont{Kennes}},
  \bibinfo{author}{\bibfnamefont{Y.}~\bibnamefont{Bai}},
  \bibinfo{author}{\bibfnamefont{B.}~\bibnamefont{Kim}}, \bibnamefont{et~al.},
  \bibinfo{journal}{arXiv preprint arXiv:1910.12147}  (\bibinfo{year}{2019}).

\bibitem[{\citenamefont{Bistritzer and MacDonald}(2011)}]{bistritzer2011moire}
\bibinfo{author}{\bibfnamefont{R.}~\bibnamefont{Bistritzer}} \bibnamefont{and}
  \bibinfo{author}{\bibfnamefont{A.~H.} \bibnamefont{MacDonald}},
  \bibinfo{journal}{Proceedings of the National Academy of Sciences}
  \textbf{\bibinfo{volume}{108}}, \bibinfo{pages}{12233}
  (\bibinfo{year}{2011}).

\bibitem[{\citenamefont{Guinea and Walet}(2018)}]{guinea2018electrostatic}
\bibinfo{author}{\bibfnamefont{F.}~\bibnamefont{Guinea}} \bibnamefont{and}
  \bibinfo{author}{\bibfnamefont{N.~R.} \bibnamefont{Walet}},
  \bibinfo{journal}{Proceedings of the National Academy of Sciences}
  \textbf{\bibinfo{volume}{115}}, \bibinfo{pages}{13174}
  (\bibinfo{year}{2018}).

\bibitem[{\citenamefont{Cea et~al.}(2019)\citenamefont{Cea, Walet, and
  Guinea}}]{cea2019electronic}
\bibinfo{author}{\bibfnamefont{T.}~\bibnamefont{Cea}},
  \bibinfo{author}{\bibfnamefont{N.~R.} \bibnamefont{Walet}}, \bibnamefont{and}
  \bibinfo{author}{\bibfnamefont{F.}~\bibnamefont{Guinea}},
  \bibinfo{journal}{Physical Review B} \textbf{\bibinfo{volume}{100}},
  \bibinfo{pages}{205113} (\bibinfo{year}{2019}).

\bibitem[{\citenamefont{Xie and MacDonald}(2020)}]{xie2020nature}
\bibinfo{author}{\bibfnamefont{M.}~\bibnamefont{Xie}} \bibnamefont{and}
  \bibinfo{author}{\bibfnamefont{A.~H.} \bibnamefont{MacDonald}},
  \bibinfo{journal}{Physical Review Letters} \textbf{\bibinfo{volume}{124}},
  \bibinfo{pages}{097601} (\bibinfo{year}{2020}).

\bibitem[{\citenamefont{Bultinck et~al.}(2019)\citenamefont{Bultinck, Khalaf,
  Liu, Chatterjee, Vishwanath, and Zaletel}}]{bultinck2019ground}
\bibinfo{author}{\bibfnamefont{N.}~\bibnamefont{Bultinck}},
  \bibinfo{author}{\bibfnamefont{E.}~\bibnamefont{Khalaf}},
  \bibinfo{author}{\bibfnamefont{S.}~\bibnamefont{Liu}},
  \bibinfo{author}{\bibfnamefont{S.}~\bibnamefont{Chatterjee}},
  \bibinfo{author}{\bibfnamefont{A.}~\bibnamefont{Vishwanath}},
  \bibnamefont{and} \bibinfo{author}{\bibfnamefont{M.~P.}
  \bibnamefont{Zaletel}}, \bibinfo{journal}{arXiv preprint arXiv:1911.02045}
  (\bibinfo{year}{2019}).

\bibitem[{\citenamefont{Liu and Dai}(2019)}]{liu2019correlated}
\bibinfo{author}{\bibfnamefont{J.}~\bibnamefont{Liu}} \bibnamefont{and}
  \bibinfo{author}{\bibfnamefont{X.}~\bibnamefont{Dai}},
  \bibinfo{journal}{arXiv preprint arXiv:1911.03760}  (\bibinfo{year}{2019}).

\bibitem[{\citenamefont{Zhang et~al.}(2020)\citenamefont{Zhang, Jiang, Wang,
  and Zhang}}]{zhang2020correlated}
\bibinfo{author}{\bibfnamefont{Y.}~\bibnamefont{Zhang}},
  \bibinfo{author}{\bibfnamefont{K.}~\bibnamefont{Jiang}},
  \bibinfo{author}{\bibfnamefont{Z.}~\bibnamefont{Wang}}, \bibnamefont{and}
  \bibinfo{author}{\bibfnamefont{F.}~\bibnamefont{Zhang}},
  \bibinfo{journal}{arXiv preprint arXiv:2001.02476}  (\bibinfo{year}{2020}).

\bibitem[{\citenamefont{Fallahazad et~al.}(2016)\citenamefont{Fallahazad,
  Movva, Kim, Larentis, Taniguchi, Watanabe, Banerjee, and
  Tutuc}}]{fallahazad2016shubnikov}
\bibinfo{author}{\bibfnamefont{B.}~\bibnamefont{Fallahazad}},
  \bibinfo{author}{\bibfnamefont{H.~C.} \bibnamefont{Movva}},
  \bibinfo{author}{\bibfnamefont{K.}~\bibnamefont{Kim}},
  \bibinfo{author}{\bibfnamefont{S.}~\bibnamefont{Larentis}},
  \bibinfo{author}{\bibfnamefont{T.}~\bibnamefont{Taniguchi}},
  \bibinfo{author}{\bibfnamefont{K.}~\bibnamefont{Watanabe}},
  \bibinfo{author}{\bibfnamefont{S.~K.} \bibnamefont{Banerjee}},
  \bibnamefont{and} \bibinfo{author}{\bibfnamefont{E.}~\bibnamefont{Tutuc}},
  \bibinfo{journal}{Physical review letters} \textbf{\bibinfo{volume}{116}},
  \bibinfo{pages}{086601} (\bibinfo{year}{2016}).

\bibitem[{\citenamefont{Rasmussen and
  Thygesen}(2015)}]{rasmussen2015computational}
\bibinfo{author}{\bibfnamefont{F.~A.} \bibnamefont{Rasmussen}}
  \bibnamefont{and} \bibinfo{author}{\bibfnamefont{K.~S.}
  \bibnamefont{Thygesen}}, \bibinfo{journal}{The Journal of Physical Chemistry
  C} \textbf{\bibinfo{volume}{119}}, \bibinfo{pages}{13169}
  (\bibinfo{year}{2015}).

\bibitem[{\citenamefont{Stier et~al.}(2016)\citenamefont{Stier, Wilson, Clark,
  Xu, and Crooker}}]{stier2016probing}
\bibinfo{author}{\bibfnamefont{A.~V.} \bibnamefont{Stier}},
  \bibinfo{author}{\bibfnamefont{N.~P.} \bibnamefont{Wilson}},
  \bibinfo{author}{\bibfnamefont{G.}~\bibnamefont{Clark}},
  \bibinfo{author}{\bibfnamefont{X.}~\bibnamefont{Xu}}, \bibnamefont{and}
  \bibinfo{author}{\bibfnamefont{S.~A.} \bibnamefont{Crooker}},
  \bibinfo{journal}{Nano letters} \textbf{\bibinfo{volume}{16}},
  \bibinfo{pages}{7054} (\bibinfo{year}{2016}).

\bibitem[{\citenamefont{Wu et~al.}(2018)\citenamefont{Wu, Lovorn, Tutuc, and
  MacDonald}}]{wu2018hubbard}
\bibinfo{author}{\bibfnamefont{F.}~\bibnamefont{Wu}},
  \bibinfo{author}{\bibfnamefont{T.}~\bibnamefont{Lovorn}},
  \bibinfo{author}{\bibfnamefont{E.}~\bibnamefont{Tutuc}}, \bibnamefont{and}
  \bibinfo{author}{\bibfnamefont{A.~H.} \bibnamefont{MacDonald}},
  \bibinfo{journal}{Physical review letters} \textbf{\bibinfo{volume}{121}},
  \bibinfo{pages}{026402} (\bibinfo{year}{2018}).

\bibitem[{\citenamefont{Zhang et~al.}(2019)\citenamefont{Zhang, Yuan, and
  Fu}}]{zhang2019moir}
\bibinfo{author}{\bibfnamefont{Y.}~\bibnamefont{Zhang}},
  \bibinfo{author}{\bibfnamefont{N.~F.} \bibnamefont{Yuan}}, \bibnamefont{and}
  \bibinfo{author}{\bibfnamefont{L.}~\bibnamefont{Fu}}, \bibinfo{journal}{arXiv
  preprint arXiv:1910.14061}  (\bibinfo{year}{2019}).

\bibitem[{\citenamefont{Kohn and Sham}(1965)}]{kohn1965self}
\bibinfo{author}{\bibfnamefont{W.}~\bibnamefont{Kohn}} \bibnamefont{and}
  \bibinfo{author}{\bibfnamefont{L.~J.} \bibnamefont{Sham}},
  \bibinfo{journal}{Physical review} \textbf{\bibinfo{volume}{140}},
  \bibinfo{pages}{A1133} (\bibinfo{year}{1965}).

\bibitem[{\citenamefont{von Barth and Hedin}(1972)}]{von1972local}
\bibinfo{author}{\bibfnamefont{U.}~\bibnamefont{von Barth}} \bibnamefont{and}
  \bibinfo{author}{\bibfnamefont{L.}~\bibnamefont{Hedin}},
  \bibinfo{journal}{Journal of Physics C: Solid State Physics}
  \textbf{\bibinfo{volume}{5}}, \bibinfo{pages}{1629} (\bibinfo{year}{1972}).

\bibitem[{\citenamefont{Rajagopal and
  Callaway}(1973)}]{rajagopal1973inhomogeneous}
\bibinfo{author}{\bibfnamefont{A.}~\bibnamefont{Rajagopal}} \bibnamefont{and}
  \bibinfo{author}{\bibfnamefont{J.}~\bibnamefont{Callaway}},
  \bibinfo{journal}{Physical Review B} \textbf{\bibinfo{volume}{7}},
  \bibinfo{pages}{1912} (\bibinfo{year}{1973}).

\bibitem[{\citenamefont{Attaccalite et~al.}(2002)\citenamefont{Attaccalite,
  Moroni, Gori-Giorgi, and Bachelet}}]{attaccalite2002correlation}
\bibinfo{author}{\bibfnamefont{C.}~\bibnamefont{Attaccalite}},
  \bibinfo{author}{\bibfnamefont{S.}~\bibnamefont{Moroni}},
  \bibinfo{author}{\bibfnamefont{P.}~\bibnamefont{Gori-Giorgi}},
  \bibnamefont{and} \bibinfo{author}{\bibfnamefont{G.~B.}
  \bibnamefont{Bachelet}}, \bibinfo{journal}{Physical review letters}
  \textbf{\bibinfo{volume}{88}}, \bibinfo{pages}{256601}
  (\bibinfo{year}{2002}).

\bibitem[{\citenamefont{Vosko et~al.}(1980)\citenamefont{Vosko, Wilk, and
  Nusair}}]{vosko1980accurate}
\bibinfo{author}{\bibfnamefont{S.~H.} \bibnamefont{Vosko}},
  \bibinfo{author}{\bibfnamefont{L.}~\bibnamefont{Wilk}}, \bibnamefont{and}
  \bibinfo{author}{\bibfnamefont{M.}~\bibnamefont{Nusair}},
  \bibinfo{journal}{Canadian Journal of physics} \textbf{\bibinfo{volume}{58}},
  \bibinfo{pages}{1200} (\bibinfo{year}{1980}).

\bibitem[{\citenamefont{Perdew and Zunger}(1981)}]{perdew1981self}
\bibinfo{author}{\bibfnamefont{J.~P.} \bibnamefont{Perdew}} \bibnamefont{and}
  \bibinfo{author}{\bibfnamefont{A.}~\bibnamefont{Zunger}},
  \bibinfo{journal}{Physical Review B} \textbf{\bibinfo{volume}{23}},
  \bibinfo{pages}{5048} (\bibinfo{year}{1981}).

\bibitem[{\citenamefont{Cole and Perdew}(1982)}]{cole1982calculated}
\bibinfo{author}{\bibfnamefont{L.~A.} \bibnamefont{Cole}} \bibnamefont{and}
  \bibinfo{author}{\bibfnamefont{J.}~\bibnamefont{Perdew}},
  \bibinfo{journal}{Physical Review A} \textbf{\bibinfo{volume}{25}},
  \bibinfo{pages}{1265} (\bibinfo{year}{1982}).

\bibitem[{\citenamefont{Perdew and Wang}(1992)}]{perdew1992accurate}
\bibinfo{author}{\bibfnamefont{J.~P.} \bibnamefont{Perdew}} \bibnamefont{and}
  \bibinfo{author}{\bibfnamefont{Y.}~\bibnamefont{Wang}},
  \bibinfo{journal}{Physical Review B} \textbf{\bibinfo{volume}{45}},
  \bibinfo{pages}{13244} (\bibinfo{year}{1992}).

\bibitem[{\citenamefont{Moruzzi et~al.}(1978)\citenamefont{Moruzzi, Janak, and
  Williams}}]{moruzzi1978computed}
\bibinfo{author}{\bibfnamefont{V.}~\bibnamefont{Moruzzi}},
  \bibinfo{author}{\bibfnamefont{J.}~\bibnamefont{Janak}}, \bibnamefont{and}
  \bibinfo{author}{\bibfnamefont{A.}~\bibnamefont{Williams}}
  (\bibinfo{year}{1978}).

\bibitem[{\citenamefont{Sticht et~al.}(1989)\citenamefont{Sticht, H{\"o}ck, and
  K{\"u}bler}}]{sticht1989non}
\bibinfo{author}{\bibfnamefont{J.}~\bibnamefont{Sticht}},
  \bibinfo{author}{\bibfnamefont{K.}~\bibnamefont{H{\"o}ck}}, \bibnamefont{and}
  \bibinfo{author}{\bibfnamefont{J.}~\bibnamefont{K{\"u}bler}},
  \bibinfo{journal}{Journal of Physics: Condensed Matter}
  \textbf{\bibinfo{volume}{1}}, \bibinfo{pages}{8155} (\bibinfo{year}{1989}).

\bibitem[{\citenamefont{Bi and Fu}(2019)}]{bi2019excitonic}
\bibinfo{author}{\bibfnamefont{Z.}~\bibnamefont{Bi}} \bibnamefont{and}
  \bibinfo{author}{\bibfnamefont{L.}~\bibnamefont{Fu}}, \bibinfo{journal}{arXiv
  preprint arXiv:1911.04493}  (\bibinfo{year}{2019}).

\bibitem[{\citenamefont{Jin et~al.}(2018)\citenamefont{Jin, Kim, Utama, Regan,
  Kleemann, Cai, Shen, Shinner, Sengupta, Watanabe et~al.}}]{jin2018imaging}
\bibinfo{author}{\bibfnamefont{C.}~\bibnamefont{Jin}},
  \bibinfo{author}{\bibfnamefont{J.}~\bibnamefont{Kim}},
  \bibinfo{author}{\bibfnamefont{M.~I.~B.} \bibnamefont{Utama}},
  \bibinfo{author}{\bibfnamefont{E.~C.} \bibnamefont{Regan}},
  \bibinfo{author}{\bibfnamefont{H.}~\bibnamefont{Kleemann}},
  \bibinfo{author}{\bibfnamefont{H.}~\bibnamefont{Cai}},
  \bibinfo{author}{\bibfnamefont{Y.}~\bibnamefont{Shen}},
  \bibinfo{author}{\bibfnamefont{M.~J.} \bibnamefont{Shinner}},
  \bibinfo{author}{\bibfnamefont{A.}~\bibnamefont{Sengupta}},
  \bibinfo{author}{\bibfnamefont{K.}~\bibnamefont{Watanabe}},
  \bibnamefont{et~al.}, \bibinfo{journal}{Science}
  \textbf{\bibinfo{volume}{360}}, \bibinfo{pages}{893} (\bibinfo{year}{2018}).

\bibitem[{\citenamefont{Huang et~al.}(2018)\citenamefont{Huang, Clark, Klein,
  MacNeill, Navarro-Moratalla, Seyler, Wilson, McGuire, Cobden, Xiao
  et~al.}}]{huang2018electrical}
\bibinfo{author}{\bibfnamefont{B.}~\bibnamefont{Huang}},
  \bibinfo{author}{\bibfnamefont{G.}~\bibnamefont{Clark}},
  \bibinfo{author}{\bibfnamefont{D.~R.} \bibnamefont{Klein}},
  \bibinfo{author}{\bibfnamefont{D.}~\bibnamefont{MacNeill}},
  \bibinfo{author}{\bibfnamefont{E.}~\bibnamefont{Navarro-Moratalla}},
  \bibinfo{author}{\bibfnamefont{K.~L.} \bibnamefont{Seyler}},
  \bibinfo{author}{\bibfnamefont{N.}~\bibnamefont{Wilson}},
  \bibinfo{author}{\bibfnamefont{M.~A.} \bibnamefont{McGuire}},
  \bibinfo{author}{\bibfnamefont{D.~H.} \bibnamefont{Cobden}},
  \bibinfo{author}{\bibfnamefont{D.}~\bibnamefont{Xiao}}, \bibnamefont{et~al.},
  \bibinfo{journal}{Nature nanotechnology} \textbf{\bibinfo{volume}{13}},
  \bibinfo{pages}{544} (\bibinfo{year}{2018}).

\bibitem[{\citenamefont{Jiang et~al.}(2018)\citenamefont{Jiang, Li, Wang, Mak,
  and Shan}}]{jiang2018controlling}
\bibinfo{author}{\bibfnamefont{S.}~\bibnamefont{Jiang}},
  \bibinfo{author}{\bibfnamefont{L.}~\bibnamefont{Li}},
  \bibinfo{author}{\bibfnamefont{Z.}~\bibnamefont{Wang}},
  \bibinfo{author}{\bibfnamefont{K.~F.} \bibnamefont{Mak}}, \bibnamefont{and}
  \bibinfo{author}{\bibfnamefont{J.}~\bibnamefont{Shan}},
  \bibinfo{journal}{Nature nanotechnology} \textbf{\bibinfo{volume}{13}},
  \bibinfo{pages}{549} (\bibinfo{year}{2018}).

\bibitem[{\citenamefont{Manchon et~al.}(2019)\citenamefont{Manchon,
  {\v{Z}}elezn{\`y}, Miron, Jungwirth, Sinova, Thiaville, Garello, and
  Gambardella}}]{manchon2019current}
\bibinfo{author}{\bibfnamefont{A.}~\bibnamefont{Manchon}},
  \bibinfo{author}{\bibfnamefont{J.}~\bibnamefont{{\v{Z}}elezn{\`y}}},
  \bibinfo{author}{\bibfnamefont{I.~M.} \bibnamefont{Miron}},
  \bibinfo{author}{\bibfnamefont{T.}~\bibnamefont{Jungwirth}},
  \bibinfo{author}{\bibfnamefont{J.}~\bibnamefont{Sinova}},
  \bibinfo{author}{\bibfnamefont{A.}~\bibnamefont{Thiaville}},
  \bibinfo{author}{\bibfnamefont{K.}~\bibnamefont{Garello}}, \bibnamefont{and}
  \bibinfo{author}{\bibfnamefont{P.}~\bibnamefont{Gambardella}},
  \bibinfo{journal}{Reviews of Modern Physics} \textbf{\bibinfo{volume}{91}},
  \bibinfo{pages}{035004} (\bibinfo{year}{2019}).

\end{thebibliography}


\end{document}